\documentclass[twocolumn,twocolappendix]{aastex631}

\usepackage{threeparttable}
\usepackage{color}
\usepackage{amsmath}
\usepackage{ulem}

\begin{document} 

\title{JWST's first glimpse of a $z > 2$ forming cluster reveals a top-heavy stellar mass function}

\author{Hanwen Sun}
\affiliation{School of Astronomy and Space Science, Nanjing University, Nanjing 210093, China}
\affiliation{Key Laboratory of Modern Astronomy and Astrophysics, Nanjing University, Ministry of Education, Nanjing 210093, China}

\author[0000-0002-2504-2421]{Tao Wang}
\affiliation{School of Astronomy and Space Science, Nanjing University, Nanjing 210093, China}
\affiliation{Key Laboratory of Modern Astronomy and Astrophysics, Nanjing University, Ministry of Education, Nanjing 210093, China}

\author{Ke Xu}
\affiliation{School of Astronomy and Space Science, Nanjing University, Nanjing 210093, China}
\affiliation{Key Laboratory of Modern Astronomy and Astrophysics, Nanjing University, Ministry of Education, Nanjing 210093, China}

\author{Emanuele Daddi}
\affiliation{AIM, CEA, CNRS, Université Paris-Saclay, Université Paris
Diderot, Sorbonne Paris Cité, F-91191 Gif-sur-Yvette, France}

\author{Qing Gu}
\affiliation{Key Laboratory for Computational Astrophysics, National Astronomical Observatories, Chinese Academy of Sciences, Beijing 100101, China}
\affiliation{Institute for Frontiers in Astronomy and Astrophysics, Beijing Normal University, Beijing
102206, China}
\affiliation{School of Astronomy and Space Science, University of Chinese Academy of Sciences, Beijing 100049, China}

\author{Tadayuki Kodama}
\affiliation{Astronomical Institute, Tohoku University, Aramaki, Aoba-ku, Sendai 980-8578, Japan}

\author{Anita Zanella}
\affiliation{Istituto Nazionale di Astrofisica (INAF), Vicolo dell’Osservatorio 5, I-35122 Padova, Italy}

\author{David Elbaz}
\affiliation{AIM, CEA, CNRS, Université Paris-Saclay, Université Paris
Diderot, Sorbonne Paris Cité, F-91191 Gif-sur-Yvette, France}

\author{Ichi Tanaka}
\affiliation{Subaru Telescope, National Astronomical Observatory of Japan, National Institutes of Natural Sciences, 650 North A’ohoku Place, Hilo, HI 96720, USA}

\author{Raphael Gobat}
\affiliation{Instituto de Física, Pontificia Universidad Católica de Valparaíso, Casilla 4059, Valparaíso, Chile}

\author{Qi Guo}
\affiliation{Key Laboratory for Computational Astrophysics, National Astronomical Observatories, Chinese Academy of Sciences, Beijing 100101, China}
\affiliation{Institute for Frontiers in Astronomy and Astrophysics, Beijing Normal University, Beijing
102206, China}
\affiliation{School of Astronomy and Space Science, University of Chinese Academy of Sciences, Beijing 100049, China}

\author{Jiaxin Han}
\affiliation{Department of Astronomy, Shanghai Jiao Tong University, Shanghai 200240, China}
\affiliation{Key Laboratory for Particle Astrophysics and Cosmology (MOE), Shanghai 200240, China}
\affiliation{Shanghai Key Laboratory for Particle Physics and Cosmology, Shanghai 200240, China}

\author{Shiying Lu}
\affiliation{School of Astronomy and Space Science, Nanjing University, Nanjing 210093, China}
\affiliation{Key Laboratory of Modern Astronomy and Astrophysics, Nanjing University, Ministry of Education, Nanjing 210093, China}
\affiliation{AIM, CEA, CNRS, Université Paris-Saclay, Université Paris
Diderot, Sorbonne Paris Cité, F-91191 Gif-sur-Yvette, France}

\author{Luwenjia Zhou}
\affiliation{School of Astronomy and Space Science, Nanjing University, Nanjing 210093, China}
\affiliation{Key Laboratory of Modern Astronomy and Astrophysics, Nanjing University, Ministry of Education, Nanjing 210093, China}

\correspondingauthor{Tao Wang}
\email{taowang@nju.edu.cn}
   
\begin{abstract}
Clusters and their progenitors (protoclusters) at $z \sim 2-4$, the peak epoch of star formation, are ideal laboratories to study the formation process of both the clusters themselves and their member galaxies. However, a complete census of their member galaxies has been challenging  due to observational difficulties. 
Here we present new JWST/NIRCam observations targeting the distant cluster CLJ1001 at $z = 2.51$ from the COSMOS-Web program, which, in combination with previous narrowband imaging targeting H$\alpha$ emitters and deep millimeter surveys of CO emitters, provide a complete view of massive galaxy assembly in CLJ1001. 
In particular, JWST reveals a population of massive, extremely red cluster members in the long-wavelength bands that were invisible in previous Hubble Space Telescope (HST)/F160W imaging (HST-dark members). Based on this highly complete spectroscopic sample of member galaxies, we show that the spatial distribution of galaxies in CLJ1001 exhibits a strong central concentration, with the central galaxy density already resembling that of low-$z$ clusters. Moreover, we reveal a ``top-heavy'' stellar mass function for the star-forming galaxies (SFGs), with an overabundance of massive SFGs piled up in the cluster core. These features strongly suggest that CLJ1001 is caught in a rapid transition, with many of its massive SFGs likely soon becoming quiescent. In the context of cluster formation, these findings suggest that the earliest clusters form from the inside out and top to bottom, with the massive galaxies in the core assembling first, followed by the less massive ones in the outskirts.

\end{abstract}



\section{Introduction}
One of the most prominent features of local galaxy clusters is a high concentration of massive quiescent galaxies (QGs) in their cores \citep{Dressler1980}. The formation process of these massive galaxies and, in particular, the role of dense environments in shaping their formation and evolution remains unclear. Galactic archeology suggests that most of their stars were formed at $z \sim 2-4$ \citep[e.g.][]{Thomas2010}, which makes (proto)clusters at these redshifts ideal laboratories to investigate their formation process. During the last decade, a number of spectroscopically confirmed (proto)clusters have been found at these redshifts \citep[e.g.][]{Pentericci2000,Kajisawa2006,Gobat2011,Hayashi2012,Overzier2016,Wang2016,Noirot2018,Oteo2018,Zhou2020,Zhou2023}. Studies of these early formed structures and their member galaxies have provided important insights into the formation process of both clusters and cluster galaxies \citep[e.g.][]{Strazzullo2016,Afanasiev2023,Mei2023,Martinez2023}. However, it is still an open question how these early formed (proto)clusters evolve into low-redshift, more mature clusters and how the dense environments impact the formation and quenching of their member galaxies. 

The major difficulty in studying galaxy formation in these high-z (proto)clusters is obtaining an unbiased census of their member galaxies. In most cases, only a small fraction of member galaxies at any given mass are spectroscopically confirmed. Without a representative and unbiased sample of cluster members, it is challenging to gauge the actual relevance of any environmental dependence. So far, various efforts have been made to obtain an unbiased census of member galaxies in a few structures at $z \gtrsim 2$, which are based on deep near-infrared (NIR) spectroscopy \citep[e.g.][]{Gobat2013,Balogh2021}, deep CO spectroscopy~\citep{Wang2018,Jin2021}, or narrowband imaging \citep{Hayashi2016,Shimakawa2018A,Shimakawa2018B,Zheng2021}. However, the prevalence of dusty starbursts and crowded distribution of galaxies in these structures often require combining several methods in order to fully and unbiasedly cover the member galaxy population. 

In this Letter, we combine the deep- and high-resolution images from JWST/NIRCam,  narrowband imaging from Subaru/MOIRCS, and CO spectroscopy from 
the Atacama Large Millimeter/submillimeter Array (ALMA)/NOEMA/Very Large Array (VLA) to provide a complete census of  member galaxies in the ${z_{{\rm{spec}}}} = 2.51$ cluster CLJ1001 (hereafter, J1001; \citealt{Wang2016}). Previous studies of J1001 have found that it has enhanced star formation rate (SFR), short gas depletion time and strong evidence that the properties of member galaxies are affected by the extreme environment \citep{Wang2016,Wang2018,Gomez-Guijarro2019,Xiao2022,Xu2023}. Here we further show a population of Hubble Space Telescope (HST)- undetected red member galaxies with bluer companions, which shows that even at $z \sim 2.5$, the high-resolution NIR-to-mid-infrared images from JWST are necessary to uncover the complete massive galaxy population of (proto)clusters. In addition, we find that J1001 has a highly concentrated density profile and a top-heavy stellar mass function (SMF), which support an inside-out and top-to-bottom evolutionary path.

The Letter is organized as follows: Section~\ref{Sec:data} presents the data set used for this work; Section~\ref{Sec:methods} shows the methods used to select member galaxies and obtain their physical properties; Section~\ref{sec: Results} presents the main results, including a population of HST-dark cluster members, the density profile, and stellar mass function.  In this work, we assume the cosmological model with ${H_0}{\rm{ = }}70~{\rm{km}} \cdot {{\rm{s}}^{ - 1}} \cdot {\rm{Mp}}{{\rm{c}}^{ - 1}}$, ${\Omega _{\rm{m}}} = 0.3$ and ${\Omega _\Lambda } = 0.7$. The AB system \citep{Oke1983} is used for the magnitudes throughout the Letter, and a \citet{Kroupa2001} initial mass function is adopted for the estimation of stellar mass.

\section{Data}\label{Sec:data}

\subsection{JWST/NIRcam images}
Cluster J1001 has been observed by JWST/NIRCam with four filters (F115W, F150W, F277W, and F444W) as part of the COSMOS-Web survey \citep{Casey2023} in December 2023. We reduce the raw JWST/NIRCam images from the Mikulski Archive for Space Telescopes (MAST)\footnote{The raw images are available at \dataset[10.17909/1r9y-dv80]{http://dx.doi.org/10.17909/1r9y-dv80}.} with the JWST Calibration Pipeline v1.12.5 \citep{Bushouse2023}. We run both stages 1 and 2 of the pipeline with default parameters. Then, we perform additional subtraction of the stripe-like 1/f noises \citep[e.g.][]{Bagley2023}, background structure, and some remaining cosmic rays \citep[e.g.][]{Rieke2023} for each ``cal" file output by stage 2. During stage 3, we use a reference catalog based on previous HST/Wide Field Camera 3 (WFC3) F160W image~\citep{Xu2023} for astrometry calibration. 

\subsection{Other multiwavelength data}
\label{Sec:multiwavelength_data}

The deep HST/WFC3 F125W and F160W images of J1001 have been obtained during Project 14750 (PI: T. Wang) with details shown in \citet{Xu2023}. These mosaics from HST have been resampled to match the pixel size of JWST/NIRCam images with Montage \citep{Berriman2003,Good2019}. In addition, to identify H$\alpha$ emitters (HAEs) within J1001, narrowband imaging with the NB2300\footnote{It is originally named as the ``CO" filter; we use the name ``NB2300" in this Letter to avoid confusion with the CO line data from ALMA.} filter on Subaru/MOIRCS has been conducted (PI: T. Kodama). The upper panel of Figure~\ref{Fig:HAEs} shows the transmission curve of the NB2300 filter. According to the FWHM of this narrowband filter, it can be used to identify HAEs at $2.483 < z < 2.519$ combined with the deep $K_{\rm s}$ imaging from UltraVista. The membership selection of J1001 with this high accuracy is essential due to the presence of a number of protoclusters at similar redshifts as J1001 in the same field~\citep{Casey2015,Cucciati2018,Huang2022}.
Based on this narrowband observation, the survey area of this work is taken as the maximum square area ($3.48' \times 3.48'$, or $1684{\rm{kpc}} \times 1684{\rm{kpc}}$) with narrowband coverage, which is shown as the blue square in Figure~\ref{Fig:RGB} (left).

CO spectroscopy in the millimeter toward J1001 has been conducted through a few programs with ALMA, NOEMA, and VLA. Most of them have been reported in previous studies~\citep{Wang2016,Wang2018,Champagne2021,Xiao2022}. New CO observations include a high-resolution CO(1-0) observation with VLA and new CO(3-2) observations with ALMA toward a larger area than that reported in ~\citet{Xiao2022}. Details of these observations will be described in a forthcoming paper (Xu et al., in preparation). Here we only use the redshift information as derived from the CO spectroscopy, totaling 23 CO(3-2) emitters including 11 CO(1-0) emitters, to select member galaxies.
 
Additionally, because J1001 is located in the COSMOS field with abundant multiwavelength data, we further retrieve archival multiband images from $U$ to IRAC channel 4 of the cluster area and perform source photometry on them. Detailed information on these images is given in Appendix~\ref{Appendix_images}.

\section{Member galaxy selection and physical properties}\label{Sec:methods}
\subsection{Catalog construction}
Based on the multiband images described in Section~\ref{Sec:data}, we build a multiwavelength photometric catalog for CLJ1001. We first convolve the F814W, F115W, F150W, and F277W images to match the point spread function (PSF) of F444W for color fidelity. Next, we perform source detection by running Source-Extractor v2.25.0 \citep{Bertin1996,Bertin2010} on the convolved F277W image, which is the deepest JWST/NIRCam image of J1001, and most sensitive to galaxies at $z \sim 2.5$. Then, we measure the total flux for each F277W-detected source in each band with Source-Extractor and T-PHOT v2.0 \citep{Merlin2015,Merlin2016}; details of the photometric methods are described in Appendix~\ref{Appendix_Photometry}.

\subsection{Sample selection}\label{sec: selection}
\begin{figure}[]
\centering
\includegraphics[width=0.48\textwidth]{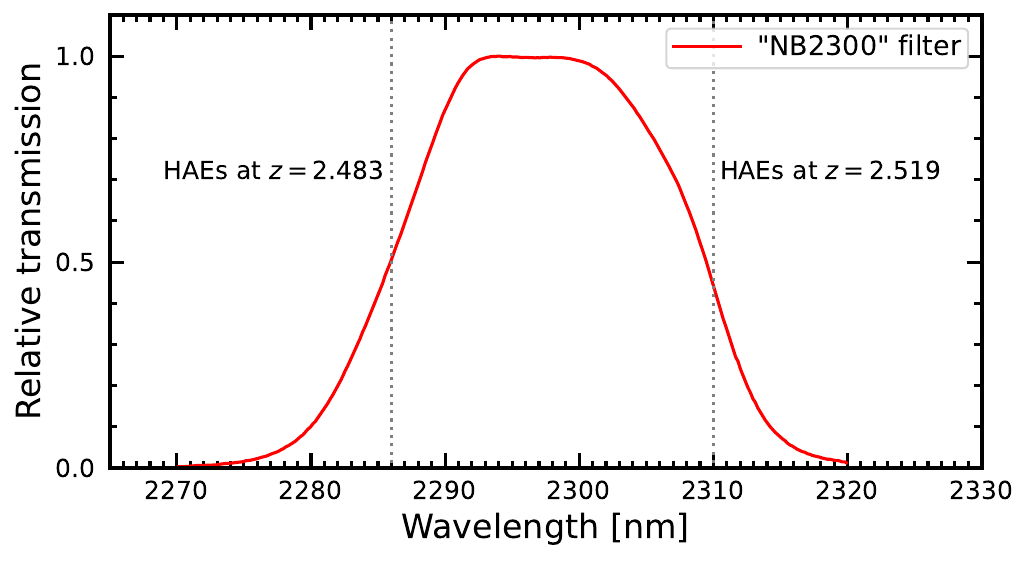}
\includegraphics[width=0.48\textwidth]{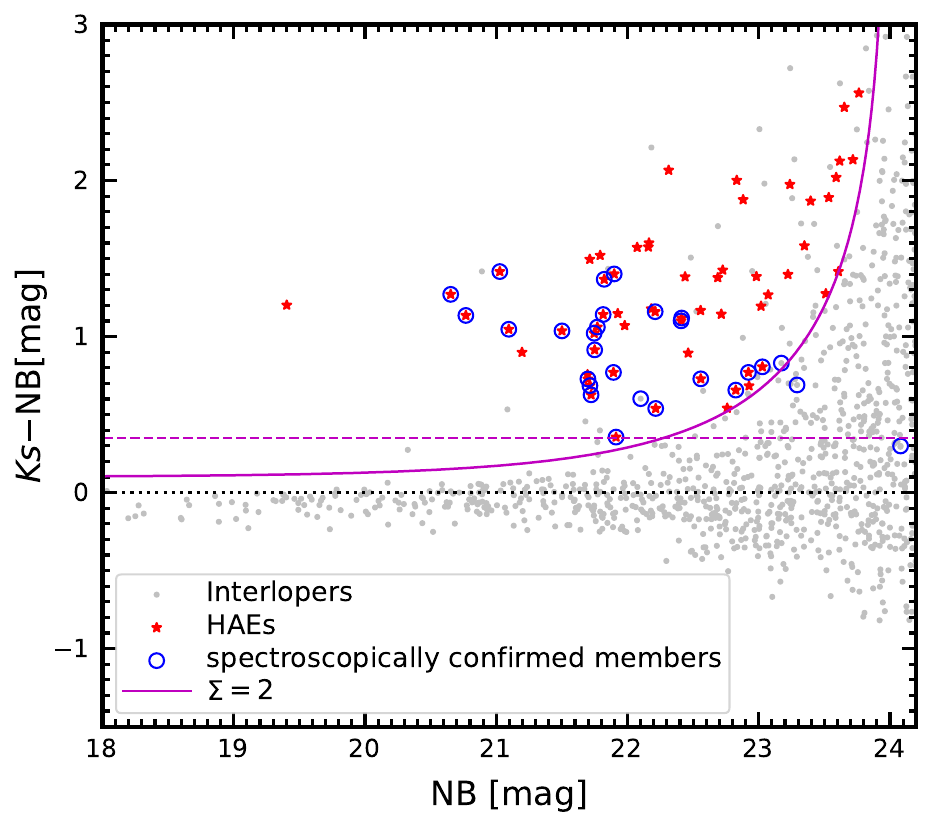}
\caption{\label{Fig:HAEs}\textbf{Upper panel}: the relative transmission curve of the NB2300 filter. The redshift range of HAEs is given by the FWHM range of this filter. \textbf{Lower panel}: the color-magnitude diagram used to select HAEs. The red stars, blue circles, and gray points are HAEs, spectroscopically confirmed members and interlopers, respectively. The magenta solid line shows the limitation given by Equation~\ref{eq_1} with $\Sigma$ = 2. The magenta dashed line corresponds to the color cut given by Equation~\ref{eq_2}.}
\end{figure}
Based on the photometric catalog, we have consistent color measurements for every F277W-detected source in the broad $K_{\rm{s}}$ and narrow NB2300 band, which are used to search for HAEs in J1001. The $K_{\rm{s}}$ and NB2300 fluxes are both measured by T-PHOT with the same configuration of the code.
Following \citet{Shimakawa2018A}, the selection criteria of HAEs are
\begin{equation}
\label{eq_1}
{K_{\rm{s}}} - {\rm{NB}} >  - 2.5\log (1 - \frac{{\Sigma \sqrt {\sigma _{Ks}^2 + \sigma _{{\rm{NB}}}^2} }}{{{f_{{\rm{NB}}}}}}) + 0.1
\end{equation}
and 
\begin{equation}
\label{eq_2}
{K_{\rm{s}}}-{\rm{NB}} > 0.35,
\end{equation}
where $f_{{\rm{NB}}}$ is the flux density in the NB2300 narrowband, ${\sigma _{Ks}}$ and ${\sigma _{\rm{NB}}}$ are the $1\sigma$ limiting flux density at the broadband ${K_{\rm{s}}}$ and narrowband NB2300, respectively; $\Sigma$ is the confidence level in $\sigma$, the color cut 0.35 \citep{Shimakawa2018A} corresponds to a rest-frame equivalent width (EW) limit of 45 \AA\ for the H$\alpha$ line; and the color term 0.1\footnote{This is a conservative value used by \citet{Hayashi2016} and is only used as a threshold for sample selection.} is applied since narrowband NB2300 is at the red end of the broadband ${K_{\rm{s}}}$, which may lead to an overestimation of the H$\alpha$ line flux for red sources. In addition, we also require that the signal-to-noise ratio (${\rm{SN}}{{\rm{R}}}$) of $f_{{\rm{NB}}}$ should be larger than 2 (${\rm{SN}}{{\rm{R}}_{{\rm{NB}}}} > 2$) for all HAE candidates. To reject emitters of other emission lines (e.g. [OIII] emitters at $z \sim 3.6$), we fit the photometric redshift ($z_{\rm phot}$) of each source using code EAZY (\citealt{Brammer:2008}; the minimum ${\chi ^2}$ redshifts are used) with agn\_blue\_sfhz\_13\footnote{\href{https://github.com/gbrammer/eazy-photoz/blob/master/templates/sfhz/README.md}{https://github.com/gbrammer/eazy-photoz/blob/master/templates/sfhz/README.md}} templates and apply 5\% systematic error floor. To calculate the uncertainty on the photometric redshifts, we compare the spectroscopic and photometric redshifts using the 28 spectroscopic members. We find two outliers with $\lvert z_{\rm phot} - z_{\rm spec} \rvert > 0.5$ (both are heavily blended), and the standard deviation of the other members is 0.043. Then, we only keep those line emitters with $2.374 < {z_{{\rm phot}}} < 2.638$, i.e., within the $3\sigma$ standard deviation of $\lvert z_{\rm phot} - z_{\rm spec} \rvert$ as HAEs. These HAEs can all be considered to have $2.483 < z < 2.519$ according to the FWHM range of the NB2300 filter.

Figure~\ref{Fig:HAEs} shows the color-magnitude diagram used to select HAEs. Here, 64 HAEs are selected by the criteria with $\Sigma > 2$, 24 of which are spectroscopically confirmed members with either H$\alpha$ detections by VLT/KMOS \citep{Wang2016} or the CO spectroscopy mentioned in section~\ref{Sec:multiwavelength_data}.
In addition to the 64 HAEs, we further include four non-HAE spectroscopic members.
All of these 68 narrowband or spectroscopic members can be selected as SFGs with their rest-frame UVJ color \citep{Williams2009,Carnall2018}. The completeness of this sample of SFGs reaches $\sim 80{\rm{\% }}$($50{\rm{\% }}$) at $\log ({M_ * }/{M_ \odot }) = 9.5(9.1)$ (see Appendix~\ref{Appendix_completeness}). Lastly, we also include six UVJ-selected quiescent members with $2.374 < {z_{{\rm{eazy}}}} < 2.638$ (three out of six quiescent members are close neighbors of confirmed members). This yields a total number of 74 cluster members. 

To derive stellar masses of these 74 member galaxies, we use the code BAGPIPES \citep{Carnall2018} within a narrow redshift range $[2.481,2.531]$ centered at $z=2.506$, with stellar population synthesis model in 2016 version of \cite{Bruzual:2003} (age$\in \rm[0.1,10]~Gyr$ and metallicity $\log Z/Z_\odot\in[0,2.5]$), a delayed star formation history (timescale $\rm \tau \in[0.3,10]~Gyr$), the dust attenuation law from \cite{Calzetti:2000} for young and old populations separately (divided by 0.01~Gyr, $A_V \in[0,5]$), and nebular emission \citep{Byler:2017} (ionization parameter $\log U\in[-5,-2]$). For members of J1001, the typical uncertainty of stellar mass is 0.06 dex, which includes the uncertainty of all other parameters and the photometric uncertainty. This has been significantly suppressed by the deep JWST/NIRCam F150W, F277W, and F444W images since they directly trace the stellar masses at $z \sim 2.5$. Two examples of the best-fit SED are shown in Appendix~\ref{Appendix_SED}.

\section{Results}\label{sec: Results}
\subsection{A population of red, massive, and HST-dark cluster members}
\begin{figure*}[htbp]
\centering
\includegraphics[width=1\textwidth]{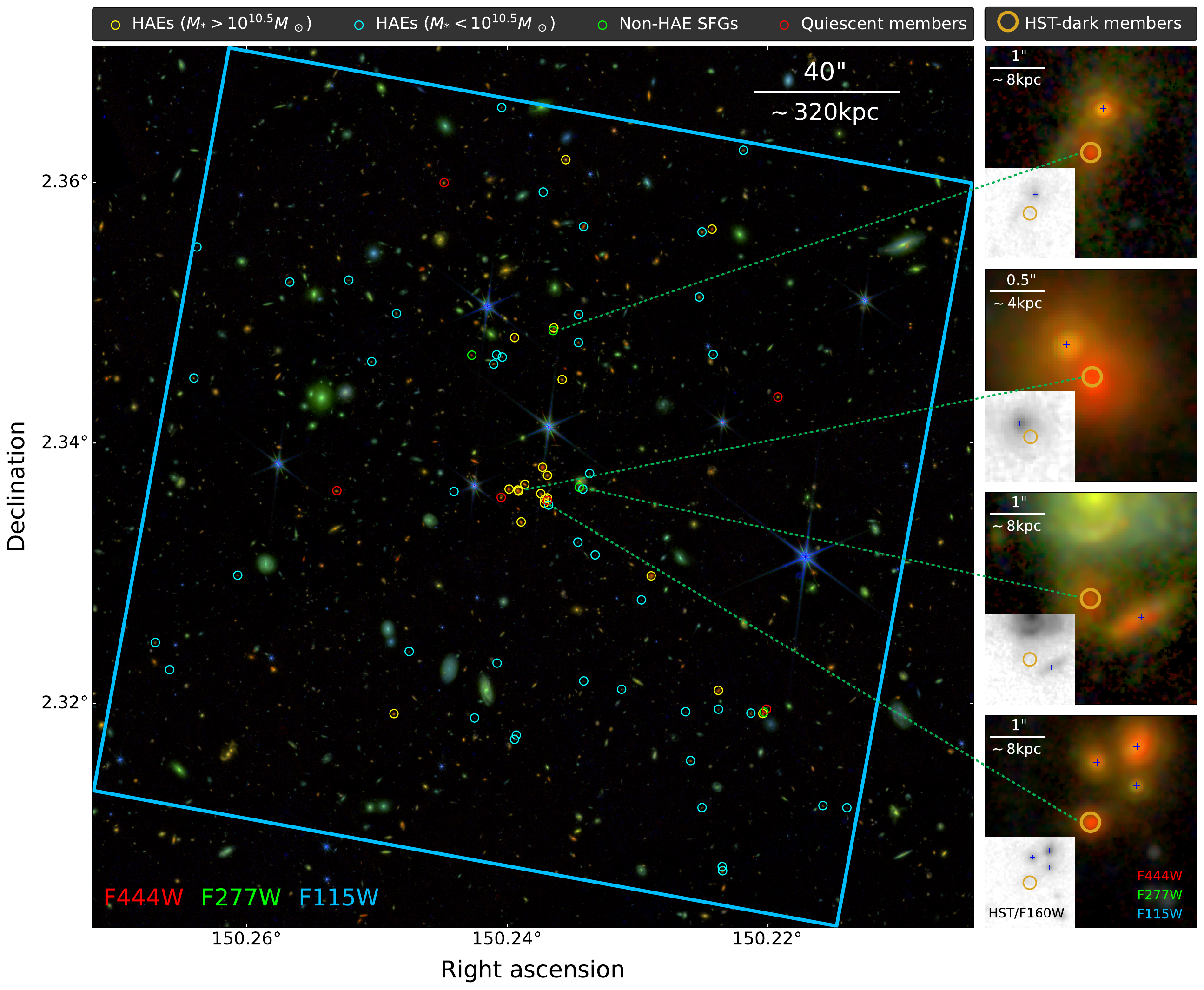}
\caption{\label{Fig:RGB}\textbf{Left}: the red, green, and blue composite color image of J1001. The R, G, and B channels of the image correspond respectively to the NIRCam/F444W, F277W, and F115W bands. The large blue square indicates the region covered by Subaru/MOIRCS narrowband imaging targeting HAEs. All member galaxies are marked with open circles, with massive HAEs, less massive HAEs, non-HAE SF members, and quiescent members denoted in yellow, cyan, green, and red, respectively. \textbf{Right}: JWST/NIRCam RGB and HST/F160W images of the four HST-dark red members (circles) and their companions (crosses).}
\end{figure*}

The spatial distribution of our selected members is shown in Figure~\ref{Fig:RGB} (left), with all member galaxies marked as open circles. 
Among these member galaxies, we find four massive red members, $\sim$ 16\% of the population at $M_{\star} > 10^{10.15} M_{\odot}$, which have been missed by previous blind source detections on the HST/WFC3 F160W image. The right panels of Figure~\ref{Fig:RGB} show the cutouts of these four HST-dark members. All of them have been spectroscopically confirmed via CO(3-2) emission by ALMA (Xu et al. in preparation).
Considering the fluxes within a small aperture ($d = 0.32"$), these HST-dark members have ${\rm{ma}}{{\rm{g}}_{{\rm{F150W,aper}}}} \sim 27$ and ${\rm{ma}}{{\rm{g}}_{{\rm{F150W,aper}}}} - {\rm{ma}}{{\rm{g}}_{{\rm{F444W,aper}}}} \sim 3$. However, their total fluxes at H band can be much higher (${\rm{ma}}{{\rm{g}}_{{\rm{F150W,total}}}} \sim 24$), which means that these HST-dark members are extremely extended at short wavelengths.
In this case, since they are often associated with one or more brighter (and bluer) companions, blind source detection on the F160W image cannot identify them and might recognize them as extended structures (e.g. tails) of their brighter companions. This suggests that the merger rate of galaxies and stellar mass function in J1001-like forming (proto)clusters can be underestimated without the JWST observation.

\subsection{A highly concentrated galaxy density profile }\label{Sec:density_profile}
\begin{figure*}[!htbp]
\centering
\includegraphics[width=0.32\textwidth]{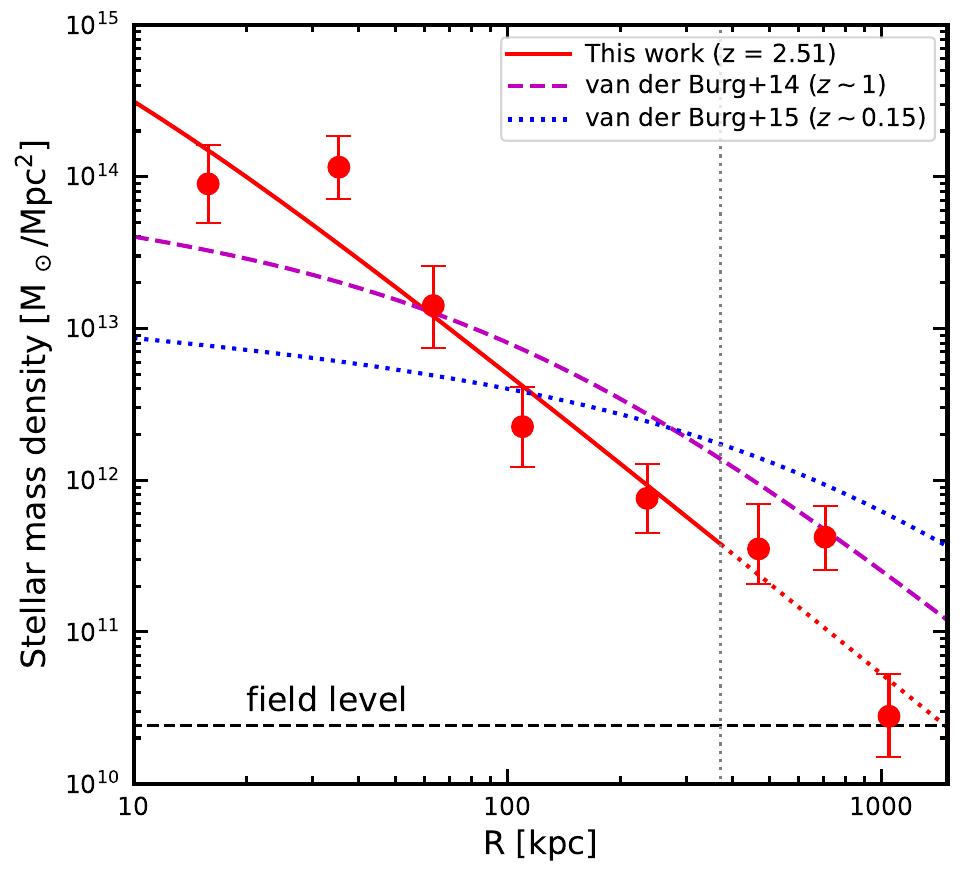}
\includegraphics[width=0.32\textwidth]{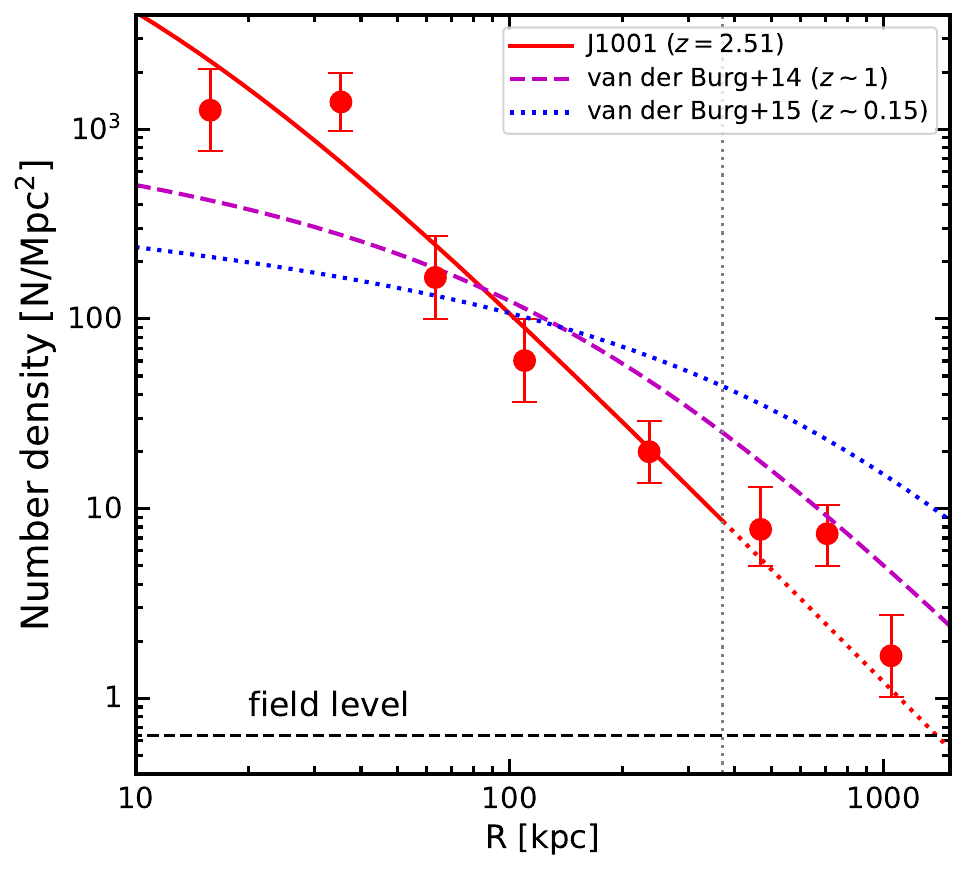}
\includegraphics[width=0.32\textwidth]{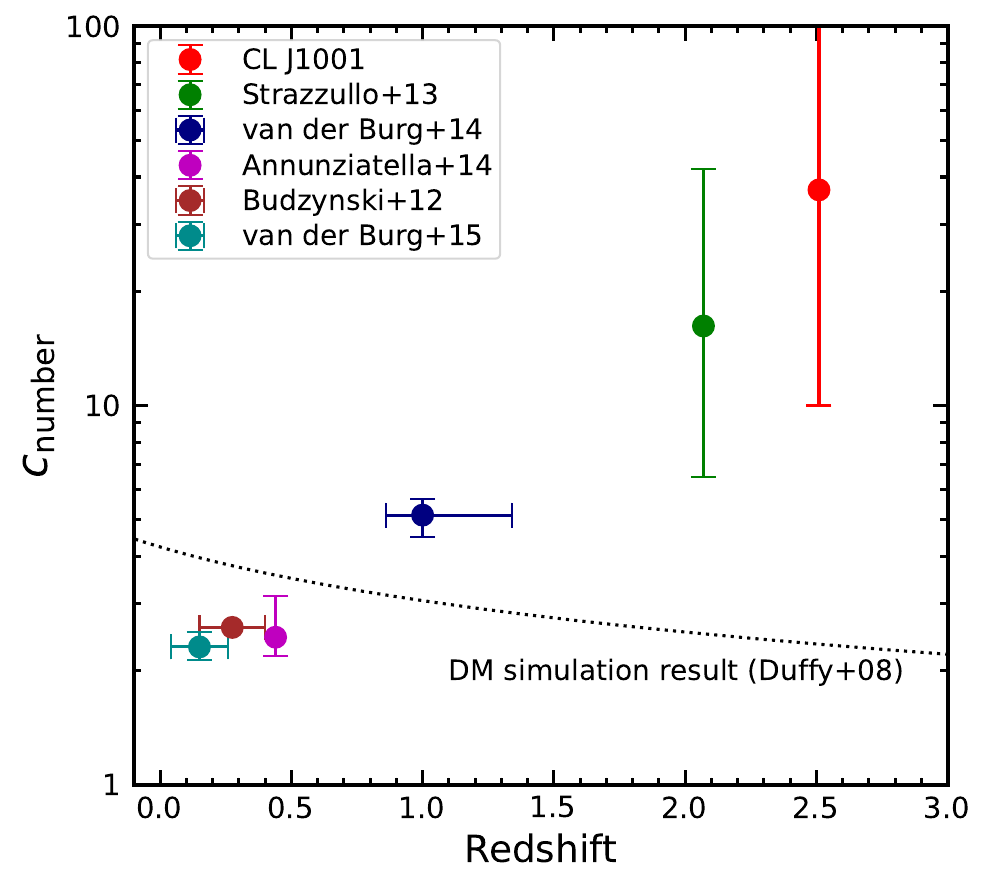}
\caption{\label{Fig:density_profile} \textbf{Left panel}: the projected stellar mass density profile of J1001 compared with the best-fit projected NFW profiles at lower redshifts given by \citet{vanderBurg2014,vanderBurg2015}. Following \citet{vanderBurg2015}, the original results with dimensionless units are converted to have physical units by assuming ${M_{200}} = 3 \times {10^{14}}{M_ \odot }$ at $z \sim 1$ and ${M_{200}} = 9 \times {10^{14}}{M_ \odot }$ at $z \sim 0.15$. Field level is given by galaxies with $2.483 < {z_{{\rm{phot}}}} < 2.519$ from the COSMOS2020 catalog \citep{Weaver2022A}. The vertical dotted line shows the ${R_{200}}$ of J1001. \textbf{Middle panel}: similar to the left panel, but the number density profiles are shown. \textbf{Right panel}: The redshift evolution of the concentration of cluster-galaxy number density profiles from observations (filled circles) and of the density profile of massive ($M_{h} = 10^{14} M_{\odot}$) dark matter halos (dotted line) based on simulations from \citet{Duffy2008}.}
\end{figure*}

We quantify the spatial distribution of member galaxies in J1001 via their projected density profile (Figure~\ref{Fig:density_profile}). To make a fair comparison with the results at $z \sim 1$ \citep{vanderBurg2014} and $z \sim 0.15$ \citep{vanderBurg2015}, we also only use galaxies with ${M_ * } > {10^{10}}{M_ \odot }$ to derive the density profile and adopt a similar IMF. Figure~\ref{Fig:density_profile} shows that the central density of J1001 can be even higher than clusters at lower redshifts, while its density in the outer region is much lower than mature clusters. According to the result from the Millennium Simulations \citep{Chiang2013}, a cluster with ${M_{200,J1001}} = {10^{13.9 \pm 0.2}}{M_ \odot }$ at $z \sim 2.5$ will have ${M_{200}} = 6.0_{ - 2.2}^{ + 3.5} \times {10^{14}}{M_ \odot }$ at $z \sim 1$. In this case, J1001 can be considered to be the predecessor of those $z \sim 1$ clusters \citep{vanderBurg2014} within the uncertainties. This, combined with the results in Figure~\ref{Fig:density_profile} supports an inside-out formation scenario \citep{vanderBurg2015} up to $z \sim 2.5$, which means the massive central galaxies in clusters are formed ahead of the smaller galaxies in the outskirts.

We then fit the data points in Figure~\ref{Fig:density_profile} with the projected Navarro-Frenk-White (NFW) profile \citep{Navarro1995,Bartelmann1996} using the ``curve\_fit" algorithm from Scipy v1.8.1 \citep{Gommers2022}. Data points outside the virial radius ${R_{200}}{\rm{ = }}369_{{\rm{ - }}53}^{{\rm{ + }}61}{\rm{kpc}}$ \citep{Wang2016} are not considered since they can be affected by subhalo structures. According to this NFW fitting, the concentration of J1001 is ${c_{\rm{mass}}}( \equiv {R_{200}}/{r_{\rm{s}}}) = 93_{ - 40}^{ + \infty }$ for stellar mass density profile and ${c_{\rm{number}}} = 37_{ - 27}^{ + \infty }$ for the number density profile, where ${\rm{ + }}\infty $ represents a single power law.

Figure~\ref{Fig:density_profile} (right) compares the ${c_{\rm{number}}}$ of J1001 with that of other clusters at lower redshifts \citep{Budzynski2012,Strazzullo2013,Annunziatella2014, vanderBurg2014,vanderBurg2015}, as well as massive dark matter halos (${M_{{\rm{halo}}}} = {10^{14}}{M_ \odot}$) from numerical simulations \citep{Duffy2008}. Previous studies of low-redshift clusters show that the concentration of member galaxy distribution decreases with time below $z \sim 1.5$ \citep{vanderBurg2015,Ahad2021}, and we further push this increasing trend of concentration with redshift for cluster galaxies to $z \sim 2.5$. These observed concentrations of galaxies can be much higher than dark matter halos because the massive galaxies (or subhalos) tend to be concentrated in the center due to dynamical friction, which makes the distribution of galaxies deviate from the host halo \citep[e.g.][]{Han2016}.  We note that the surveys shown here have different fitting ranges and virial radii, which may also affect the concentration, and we do not compare ${c_{\rm{mass}}}$ since it is too sensitive to the presence of massive central galaxies.

\subsection{A ``Top-heavy'' stellar mass function of star-forming member galaxies}

\begin{figure}[]
\centering
\includegraphics[width=0.5\textwidth]{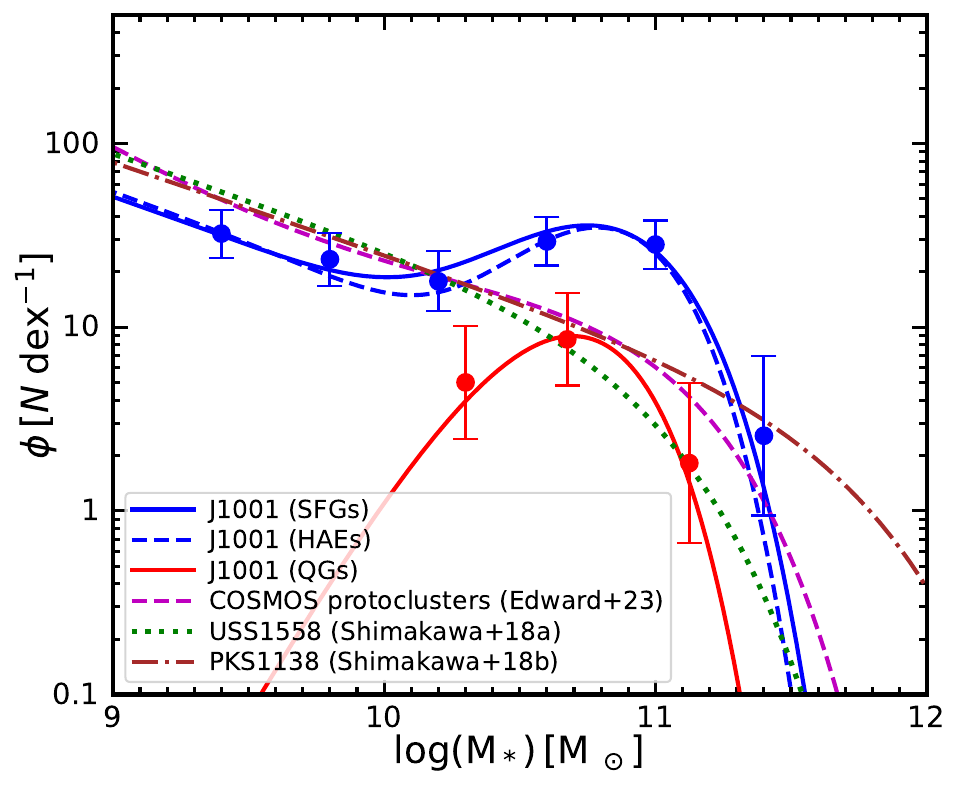}
\caption{\label{Fig:SMF} SMF of cluster J1001 compared with results in other protoclusters at $z \sim 2$. The blue and red solid lines show the SMF of SFGs and QGs in cluster J1001 at $z = 2.506$. The blue dashed line is the SMF of HAEs in J1001, which provides fair comparisons with \citet{Shimakawa2018A,Shimakawa2018B}. The purple dashed line is the stacked SMF of SFGs in 14 (proto)clusters between $z = 2.0 - 2.5$ in the COSMOS field \citep{Edward2023}, the green dotted line is the SMF of HAEs in cluster USS 1558 at $z = 2.5$ \citep{Shimakawa2018A}, the brown dashed-dotted line is the SMF of HAEs in cluster PKS 1138 at $z = 2.2$ \citep{Shimakawa2018B}. All the three SMFs of other (proto)clusters have been renormalized to match the normalization of J1001.}
\end{figure}

\begin{figure*}[]
\centering
\includegraphics[width=1\textwidth]{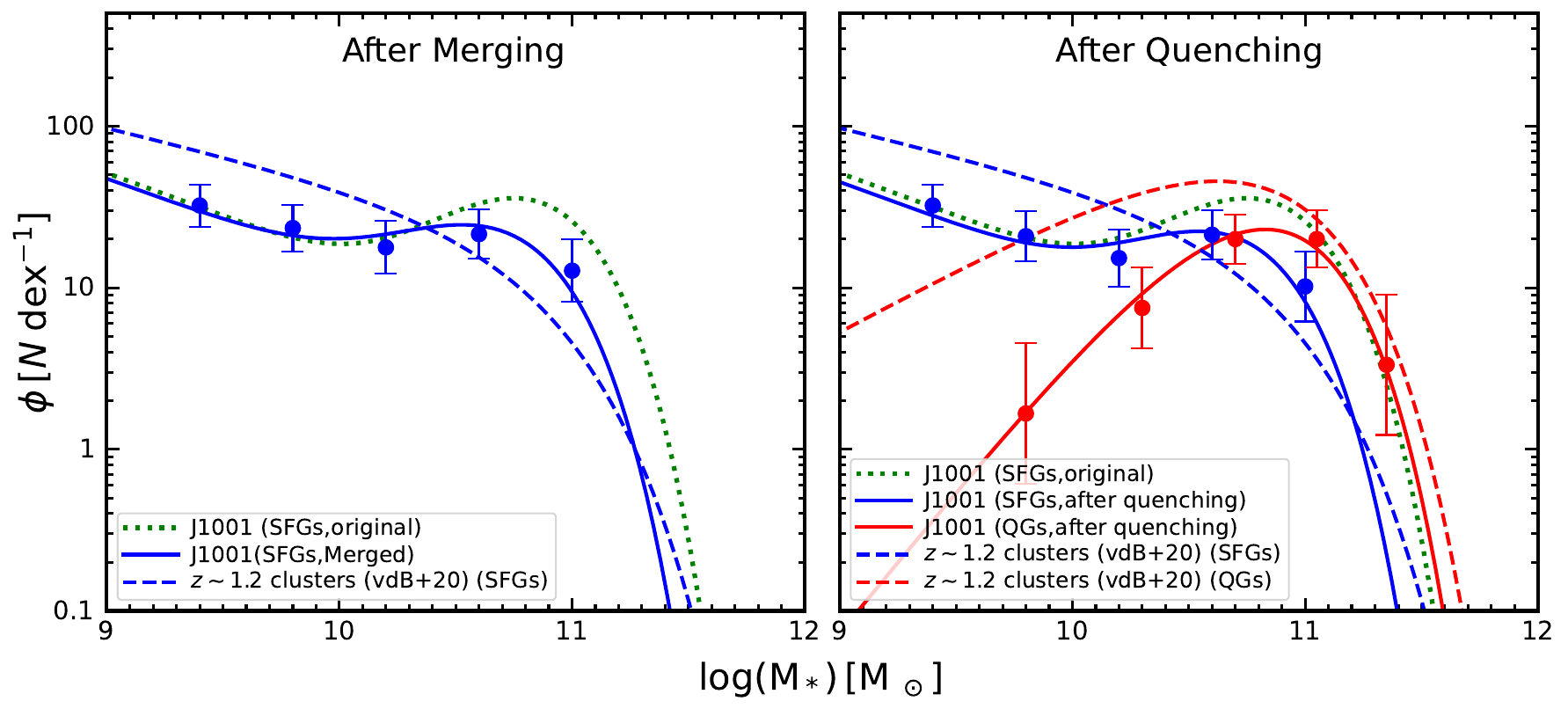}
\caption{\label{Fig:SMF_test} \textbf{Left panel}: similar to Figure~\ref{Fig:SMF}, while all the massive central galaxies within 50 kpc are merged into a quiescent BCG. The green dotted line is a copy of the blue line in Figure~\ref{Fig:SMF}. The dashed line is the SMF of clusters at $z \sim 1.2$ \citep{vanderBurg2020} within 1Mpc of their center (comparable with the size of our survey). \textbf{Right panel}: the SF and quiescent SMF of cluster J1001 with all galaxies to be quenched within 0.5Gyr considered as QGs. The blue and red dashed lines show the SF and quiescent SMF of cluster galaxies at $z \sim 1.2$, respectively.}
\end{figure*}

The combination of deep JWST/NIRCam imaging, Subaru/MOIRCS narrowband imaging, and ALMA/NOEMA/VLA CO line surveys yields a complete census of star-forming (SF) members at ${M_ *} \gtrsim {10^{9.5}}{M_ \odot }$ (see Appendix~\ref{Appendix_completeness}) in J1001. Based on this sample, we determine the SMF of SF members in J1001 (Figure~\ref{Fig:SMF}). Instead of using the data points in Figure~\ref{Fig:SMF}, we perform the maximum-likelihood fitting to avoid the arbitrary binning procedure. We fit the SMF with both the single Schechter function
\begin{equation}
\label{eq_3s}
\begin{aligned}
\Phi {\rm{d}}(\log M) = & \ln (10) \times \exp ( - {10^{\log M - \log {M^ * }}}) \\
& \times [\Phi _1^ * {({10^{\log M - \log {M^ * }}})^{{\alpha _1} + 1}}]{\rm{d}}(\log M),
\end{aligned}
\end{equation}
which is a single gamma distribution of the term ${10^{\log M - \log {M^ * }}}$, and also the double Schechter function
\begin{equation}
\label{eq_3}
\begin{aligned}
\Phi {\rm{d}}(\log M) = & \ln (10) \times \exp ( - {10^{\log M - \log {M^ * }}}) \\
& \times [\Phi _1^ * {({10^{\log M - \log {M^ * }}})^{{\alpha _1} + 1}}\\
 & + \Phi _2^ * {({10^{\log M - \log {M^ * }}})^{{\alpha _2} + 1}}]{\rm{d}}(\log M),
\end{aligned}
\end{equation}
which is the mixture of two single Schechter functions and is similar to the mixtures of gammas discussed by \citet{Young2019}. Then, we determine which model is better with the Bayesian Information Criterion (BIC) and find $\Delta ({\rm{BIC}}) \equiv {\rm{BI}}{{\rm{C}}_{{\rm{single}}}} - {\rm{BI}}{{\rm{C}}_{{\rm{double}}}} = 11.02$. This is above the typical threshold for a valid comparison $\Delta ({\rm{BIC}}) > 8 \sim 10$ \citep{Kass1995}, which means that the SMF of SFGs significantly requires a second Schechter component.
Meanwhile, we also plot the SMF of QGs in J1001, which can be well fitted by a single Schechter function with a negative $\Delta ({\rm{BIC}}) =  - 2.21$. Details of the fitting method and best-fit parameters are shown in Appendix~\ref{Appendix_fitting_results}.

Compared with the SMF shape of other protoclusters \citep{Shimakawa2018A,Shimakawa2018B,Edward2023},
the SMF of SFGs within J1001 shows a clear top-heavy feature around ${M_*} \sim {10^{10.5 - 11.0}}{M_ \odot }$. This means that J1001 has an excess of massive SFGs, most of which are concentrated in the cluster core (Figure~\ref{Fig:RGB}). This is consistent with the high concentration of J1001 shown in Figure~\ref{Fig:density_profile}. Low-redshift clusters only exhibit a top-heavy structure when both the QGs and SF members are combined \citep[e.g.][]{vanderBurg2013,vanderBurg2018,vanderBurg2020,Annunziatella2014}. A direct indication is that the massive SF members in J1001, at the time of the observation, likely include a large population of progenitors of massive QGs. Another possibility is that many of these massive SFGs may merge into a single brightest cluster galaxy (BCG), considering that many of them are located within a small volume. 

We illustrate these two scenarios on the possible evolutionary path of the SMF in J1001 in Figure~\ref{Fig:SMF_test}. Firstly, consider that most of the massive SFGs in J1001 in the central area might soon be merged into a BCG, which is currently still missing for J1001. Specifically, we assume all galaxies within 50 kpc of the cluster center will be merged into a quiescent BCG, which can then be excluded from the SMF of SFGs. In this case, as shown in the left panel of Figure~\ref{Fig:SMF_test}, the top-heavy feature will be obviously suppressed with
$\Delta ({\rm{BIC}})$ decreasing from 11.02 to 1.01.  

Secondly, we consider the effect of quenching on the evolution of the SMF of both SFGs and QGs. We collect the gas depletion time ${t_{{\rm{dep}}}} \equiv {M_{{\rm{gas}}}}/{\rm{SFR}}$ for members with CO spectroscopy (see section~\ref{Sec:multiwavelength_data}) and assume that the SFGs with ${t_{{\rm{dep}}}} < 0.5{\rm{Gyr}}$ will soon become QGs. After this transformation, we  derive the yielding SMF of SFGs and QGs in J1001, as shown in the right panel of Figure~\ref{Fig:SMF_test}. We find that the top-heavy feature in the original SMF of SFGs in J1001 is strongly suppressed with $\Delta ({\rm{BIC}})$ decreasing to 0.96. Meanwhile, a rapid rise for the SMF of QGs in J1001 is observed, which surpasses that of SFGs at the massive end and matches QGs in low-redshift clusters. 

The above analysis shows that both mergers and quenching can help eliminate the ``top-heavy'' feature in the SMF of SFGs in J1001. In practice, both processes might take place simultaneously. Moreover, by comparison with the SMF in $z \sim 1.2$ clusters from the GOGREEN program~\citep{vanderBurg2020}, we show that the massive end of J1001 is already as abundant as the clusters at $z \sim 1.2$, while the number of low-mass members with $\log {M_{\rm{*}}} = 9 - 10$ is much smaller. We have briefly discussed that J1001 can be approximately considered as the predecessor of clusters observed at $z\sim1$ (section~\ref{Sec:density_profile}). In addition, the dearth of small members in J1001 is unlikely to be an observational effect since our sample should be complete at $\log {M_{\rm{*}}} > 9.5$ (Appendix~\ref{Appendix_completeness}), and all the expected incompleteness has already been corrected. This means that the massive cluster galaxies are already in place at $z \sim 2.5$, suggesting a top-to-bottom growth of these early formed clusters.

\section{Discussion and Conclusions}\label{sec: Conclusions}
Based on the deep multiwavelength observations including JWST/NIRCam and narrowband imaging with Subaru/MOIRCS, we have performed a complete census of the member galaxies in cluster J1001 down to $M_{\star} \gtrsim 10^{9} M_{\odot}$ at $z=2.51$. In this Letter, we focus on the global properties of this sample of member galaxies, including their density profile and stellar mass function, aiming to constrain the evolutionary state of J1001 in the context of cluster formation at high redshifts. Our main findings are listed as follows:

1. JWST/NIRCam observations reveal a population of red and massive cluster members, which have been missed from previous deep HST/F160W imaging (HST-dark). In addition to their faintness, most of them have a close, bluer HST companion. This inhibits blind detections of the red sources in HST images, which only become detectable at longer wavelength by JWST/NIRCam with better sensitivity and resolution. The prevalence of the massive, HST-dark cluster members suggests that JWST is necessary to obtain a complete census of member galaxies, even at $z \sim 2.5$. Consequently, previous estimates based on HST or ground-based observations on the massive end of the mass function, as well as the merger rate of member galaxies, might be underestimated.

2. We find that the spatial distribution of member galaxies in J1001 is highly concentrated. By performing NFW profiling fitting of the stellar mass and number density profiles, we extend previous measurements on the concentration of cluster member galaxy distribution to $z \sim 2.5$. In contrast to the cosmic evolution of dark matter halos, we find that cluster galaxy distribution at higher redshifts exhibits a higher concentration. The central stellar density of J1001 is already comparable to or even higher than more massive low-redshift clusters, while the outskirts lie below that of low-redshift clusters. This strongly suggests an inside-out formation scenario, at least for these early formed clusters.

3. Based on the mass-complete sample of star-forming members, we show that the stellar mass function of J1001 shows a prominent ``top-heavy'' feature, with overabundant massive SF members compared to a single Schechter function. The total number and stellar density of these massive SFG members are comparable to that of massive SFG and QGs combined in clusters at lower redshifts. Together with the low quiescent fraction in J1001, these findings suggest a minimal role of preprocessing, and most of the QGs should be quenched only after they were accreted onto the cluster.

These findings provide novel insights into the formation processes of clusters and their member galaxies at $z \sim 2.5$. Based on a complete census of member galaxies, we show that the galaxy density profile and stellar mass functions are powerful tools to characterize the evolutionary state of galaxy (proto)clusters.  
The centrally concentrated galaxy density profile and the ``top-heavy'' stellar mass function of the SF members in J1001 indicate that these early formed clusters and their member galaxies grow in an inside-out and top-bottom fashion: most of the massive galaxies in the core are assembled first, and less massive galaxies in the outskirts are formed later. Future studies with much larger samples of (proto)clusters at $z \sim 2-4$ will provide a more comprehensive understanding of clusters and their member galaxy formation at their peak formation epoch \citep{Zhou2023}.

\section*{Acknowledgments}
We thank the anonymous referee and Dr. Veronica Strazzullo for the helpful comments on this Letter. T.W. acknowledges support by National Natural Science Foundation
of China (Project No. 12173017 and Key Project No. 12141301),
National Key R\&D Program of China (2023YFA1605600), and the
China Manned Space Project (No. CMS-CSST-2021-A07). J.H. acknowledges support from the China Manned Space Project (No. CMS-CSST-2021-A03) and the Yangyang Development Fund.

Facilities: JWST, HST, Subaru, Vista, Spitzer, CFHT

\appendix
\restartappendixnumbering
\section{Archival images}\label{Appendix_images}
In this work, the archival multiband images from $U$ to IRAC channel 4 of the COSMOS field are used, including the $U$ band from The COSMOS-WIRCam Near-Infrared Imaging Survey taken with the Canada–France–Hawaii Telescope (CFHT) \citep{McCracken2010}; $B$ and $IB427$ from the COSMOS-20 survey taken with Subaru/Suprime-Cam \citep{Taniguchi2015}; $g$, $r$, $i$, and $z$ from the third public data release (PDR3) of the Hyper Suprime-Cam Subaru Strategic Program (Subaru/HSC-SSP) \citep{Aihara2022}; the F814W band from the Cosmic Evolution Survey taken with the Advanced Camera for Surveys on HST (HST/ACS; \citealt{Scoville2007}); $Y$, $J$, $H$, and ${K_{\rm{s}}}$ band from the fourth data release of the UltraVISTA near-infrared imaging survey \citep{McCracken2012}; an archival HST/WFC3 F110W image from \citet{Negrello2014}; and the Spitzer/IRAC channel 1, 3, and 4 images from the Spitzer Large Area Survey with Hyper-Suprime-Cam (SPLASH) \citep{Steinhardt2014}.

\section{The methods of photometry}\label{Appendix_Photometry}
During the construction of our multiwavelength photometric catalog, we combine different methods of photometry for different bands since our data cover a wide resolution range. Firstly, for the high-resolution images from HST/ACS and JWST, we run Source-Extractor v2.25.0 \citep{Bertin1996,Bertin2010} in the dual image mode to obtain the Kron aperture photometry (\citealt{Kron1980}; FLUX\_AUTO from Source-Extractor), during which the fluxes are measured on the convolved images with the resolution of the JWST/NIRCam F444W image. For the HST/WFC3 images whose resolutions are slightly larger than the F444W image, we get Kron aperture photometry without performing a PSF match on them. Instead, we further convolve the F150W image to match the PSF size of HST/WFC3 images and take the ratio of the F150W fluxes after and before this convolution ${f_{150,{\rm{F444W\thinspace resolution}}}}/{f_{150,{\rm{WFC3\thinspace resolution}}}}$ as the correction factor for WFC3 fluxes. We note that these direct results given by Kron aperture photometry can miss some fluxes in the outskirts of each source, especially since a nonnegligible fraction ($\sim 10\%$) of light can be scattered to large radii by the PSF of JWST/NIRCam. In this case, to obtain the accurate total flux, we estimate the ratio of missed light by performing Kron aperture photometry on the PSF and correct the Kron fluxes based on this ratio.

For other images with poorer resolution than HST and JWST, we use T-PHOT v2.0 \citep{Merlin2015,Merlin2016} to obtain prior-based deblended photometry based on the source positions and morphologies detected in the JWST/NIRCam images. T-PHOT is a deconfusion photometry software that convolves the cutouts of each source from a high-resolution image to produce low-resolution priors and then fits the flux of the same source in a low-resolution image with these priors. For images from ground-based telescopes and Spitzer, we perform T-PHOT photometry using the image of the closest JWST/NIRCam band as high resolution prior to obtaining deblended total flux for each F277W-detected source. 

\begin{figure}[htbp]
\centering
\includegraphics[width=0.48\textwidth]{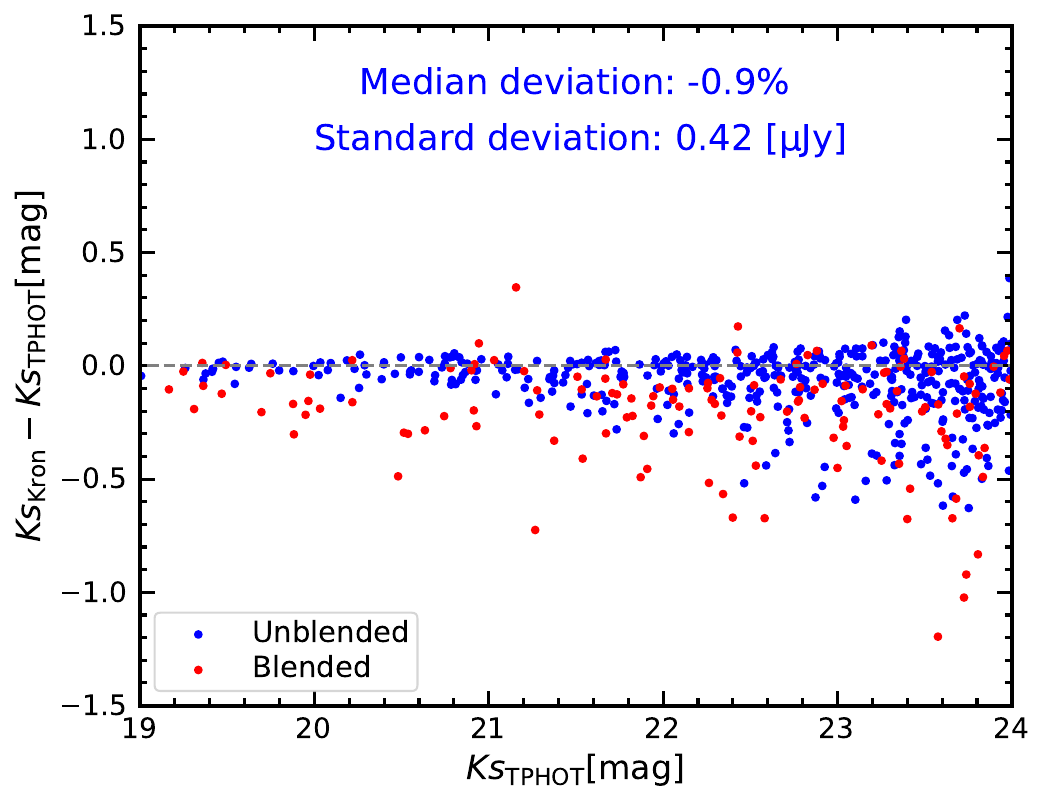}
\caption{\label{Fig:Ks_cross} The comparison of the $K_{\rm{s}}$-band fluxes measured with T-PHOT and the Kron method. The blue and red points show the unblended and blended sources, respectively.}
\end{figure}

When combining the results measured by different methods, \citet{Merlin2021} have shown that the results of T-PHOT photometry do not have systematic bias compared with Kron aperture photometry. In addition, we also perform a similar test with our data: We directly perform source detection and Kron aperture photometry on the $K_{\rm{s}}$-band image from UltraVISTA \citep{McCracken2012} and cross-match our F277W-detected catalog with this blind $K_{\rm{s}}$-band catalog. Figure~\ref{Fig:Ks_cross} shows the comparison of the $K_{\rm{s}}$ fluxes measured by T-PHOT and Kron apertures. We find that the two fluxes are consistent for the unblended sources with a median deviation $-0.9\%$. Meanwhile, the T-PHOT fluxes of blended sources can be much smaller and more reliable since the Kron results can be contaminated by nearby sources. 

\section{Examples of the best-fit SED}\label{Appendix_SED}
\begin{figure*}[htbp]
\centering
\includegraphics[width=0.75\textwidth]{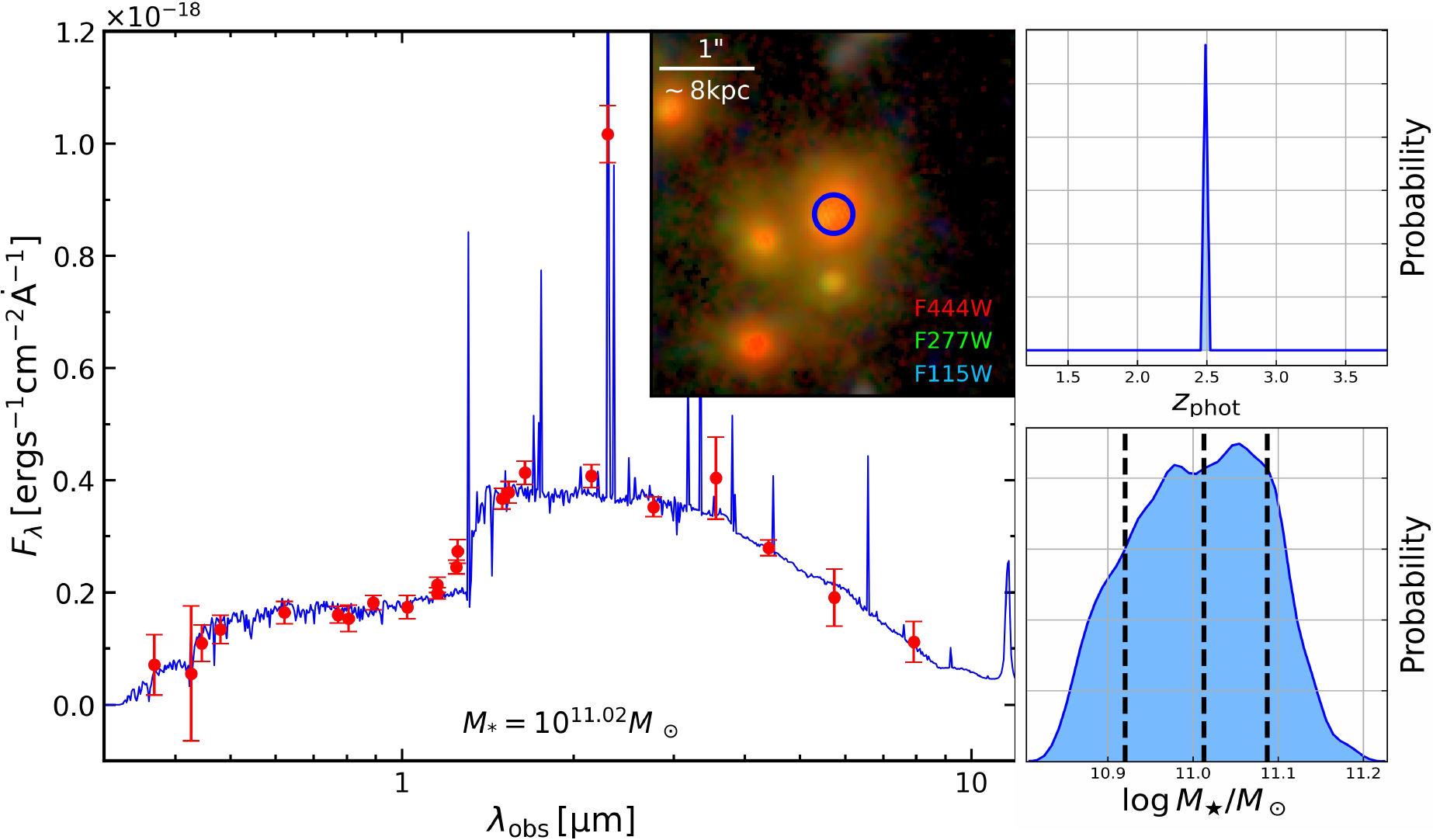}
\includegraphics[width=0.75\textwidth]{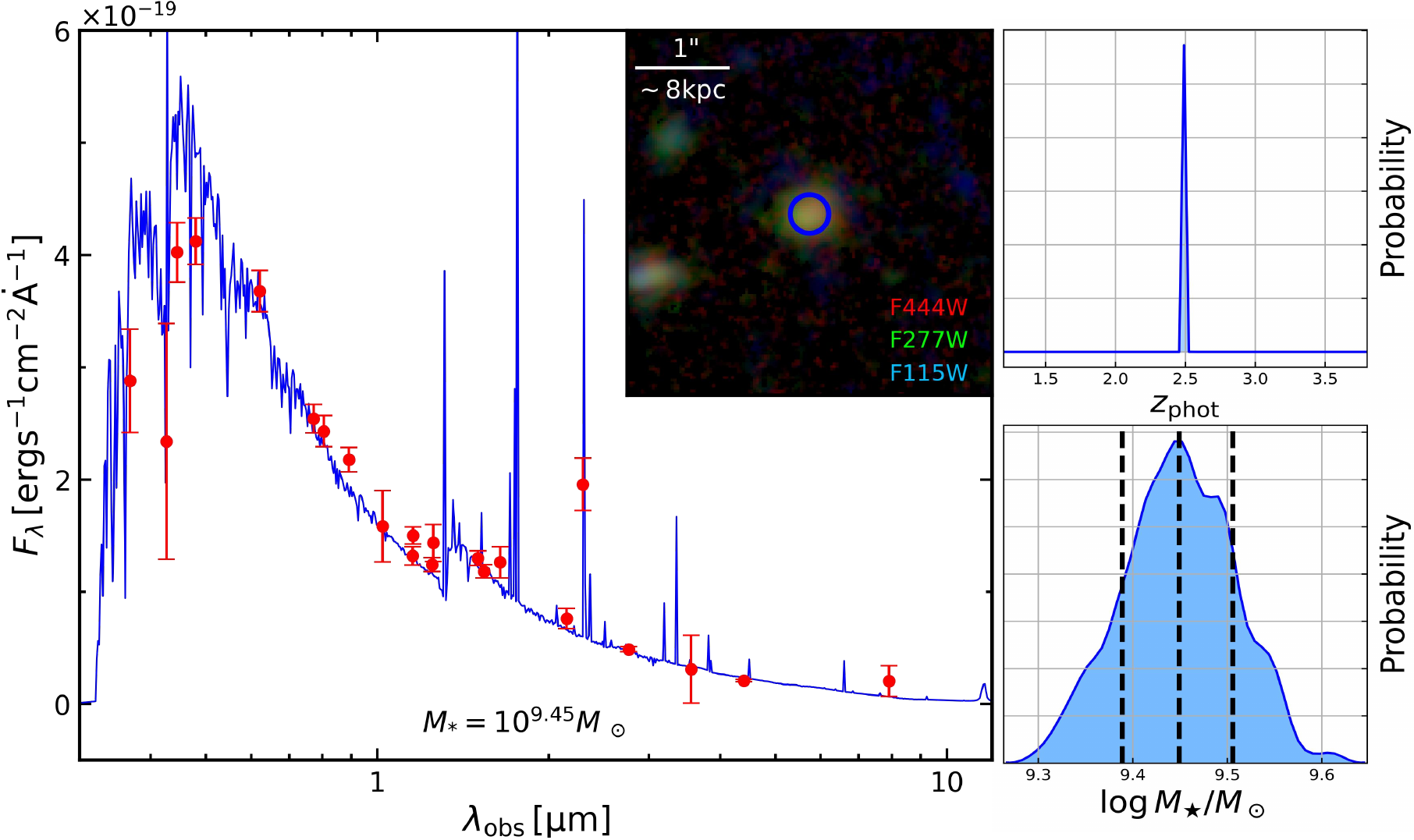}
\caption{\label{Fig:SED}\textbf{Left panels}: examples of the best-fit SED of two member galaxies from Bagpipes. \textbf{Right panels}: the probability density distribution functions of the photometric redshift from EAZY and stellar mass from Bagpipes.}
\end{figure*}

Two examples of the best-fit SED given by Bagpipes are shown in Figure~\ref{Fig:SED}. The first one is representative of massive members that dominate the top-heavy feature, while the second one is representative of small members around the limiting depth of this work.
 
\section{Sample completeness}\label{Appendix_completeness}
\begin{figure}[!htb]
\centering
\includegraphics[width=0.45\textwidth]{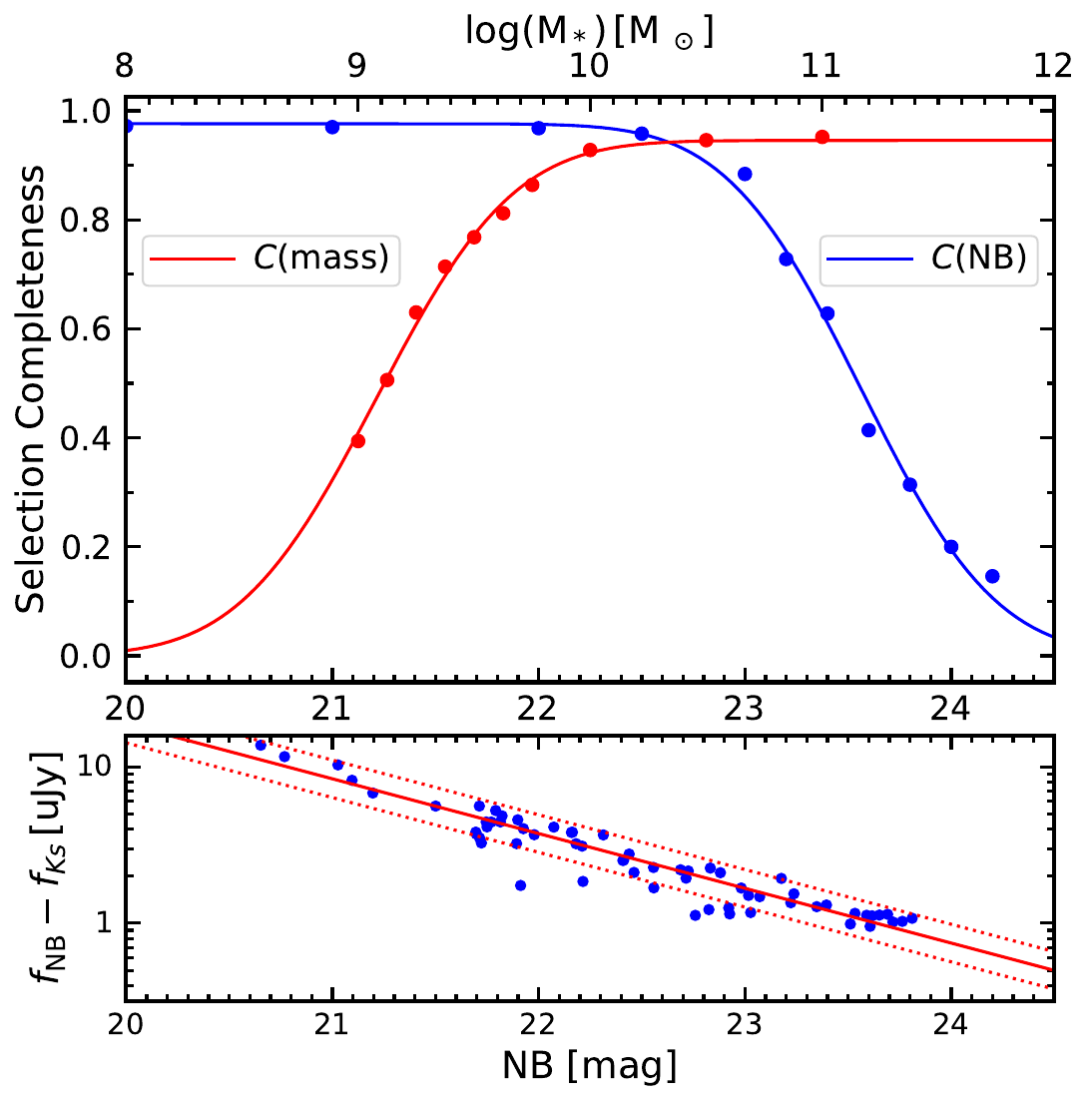}
\caption{\label{Fig:SC}\textbf{Upper panel}: the selection completeness. Blue and red points are the simulated selection completeness as a function of narrowband magnitude and stellar mass, respectively. And the blue and red lines are the completeness function fitted by the complementary error function. \textbf{Lower panel}: the narrowband exceeds fluxes as a function of narrowband magnitude. The red solid line is the result of linear fitting, while the dotted lines show the scatter of 0.12 dex.}
\end{figure}
We present the derivation of selection completeness with Monte Carlo simulation in this appendix. Firstly, following \citet{Shimakawa2018B}, selection completeness is considered as a function of narrowband magnitude. At each given narrowband magnitude, we embed 500 point sources into the real narrowband and ${K_{\rm{s}}}$-band image. The ${K_{\rm{s}}}$ fluxes of these sources are determined by their narrowband magnitude according to the best-fit power-law relation shown in the lower panel of Figure~\ref{Fig:SC} with a random 0.12 dex scatter. This tight relation between the flux excess and narrowband magnitude is fitted with the 64 real HAEs from observation. Next, we perform T-PHOT photometry on the simulated narrowband and ${K_{\rm{s}}}$-band images, during which the simulated point sources are also embedded into the F277W image to provide the high-resolution priors. We then test the recovery rate of the membership selection procedure described in Section~\ref{sec: selection}, and fit the resulting selection completeness with a complementary error function. The best-fit result is
\begin{equation}
\label{eq_A1}
C{({\rm{NB}})} = 0.488 \times {\rm{Erfc[1}}{\rm{.372}} \times {\rm{(NB}} - {\rm{23}}{\rm{.565)]}}
\end{equation}
This best-fit error function used for completeness correction is shown in Figure~\ref{Fig:SC} (upper).

\begin{figure}[htbp]
\centering
\includegraphics[width=0.45\textwidth]{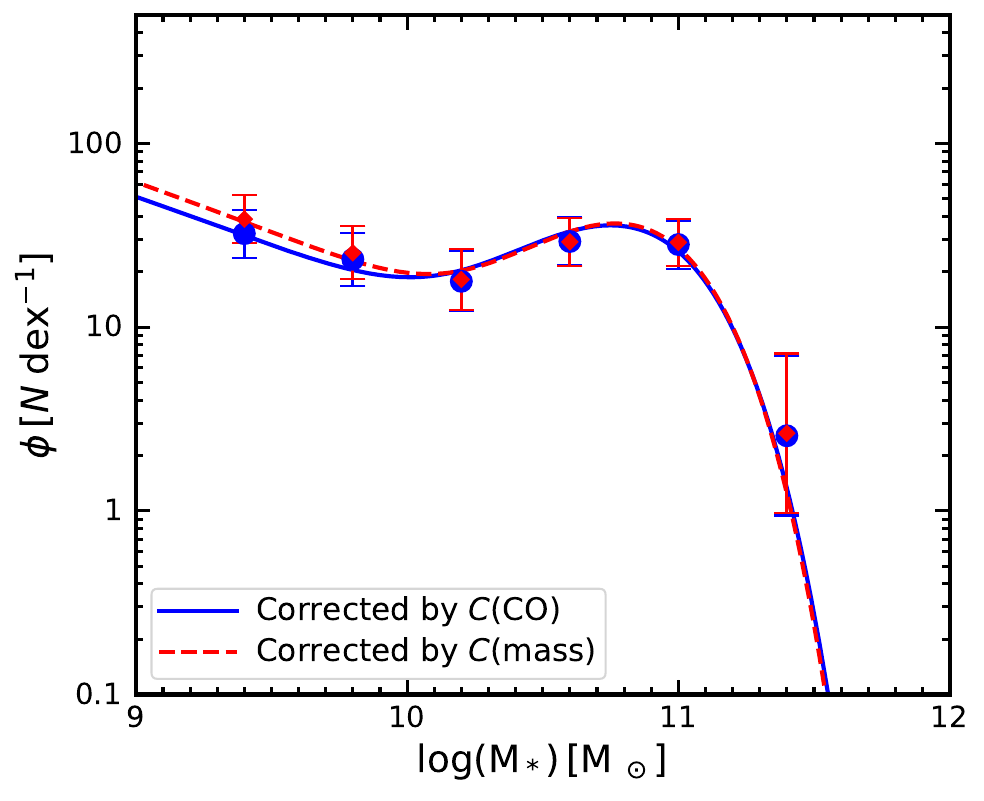}
\caption{\label{Fig:SMF_2CC}Comparison of the SMF under two different incompleteness-correction procedures. The blue solid line is the SMF corrected by Equation~\ref{eq_A1} (identical with the blue line in Figure~\ref{Fig:SMF}). The red dashed line is the SMF corrected by Equation~\ref{eq_A3}.}
\end{figure}

Next, we also test the selection completeness as a function of stellar mass. To determine the ${K_{\rm{s}}}$\-band fluxes for sources at a given mass, we collect the galaxies at $2.4 < z < 2.6$ within the ultradeep stripes of the UltraVISTA survey \citep{McCracken2012} from the COSMOS2020 catalog \citep{Weaver2022A}, and fit the stellar mass-${K_{\rm{s}}}$ flux relation at each given mass with a Gaussian distribution. Then, the ${K_{\rm{s}}}$-band fluxes of simulated sources can be randomly generated using this best-fit Gaussian distribution. As for narrowband fluxes, we firstly generate the SFR for each source based on its stellar mass by taking the SF main sequence at $z = 2.5$ \citep{Popesso2023} with a 0.3 dex scatter. This SFR can be converted to a H$\alpha$ flux based on the Equation 2 of \citet{Kennicutt1998} with a reduction of 0.24 dex for IMF calibration. Since the redshift of J1001 is close to the edge given by the NB2300 filter (Figure~\ref{Fig:HAEs}), we randomly give the redshift of each simulated source using a Gaussian distribution given by fitting the distribution of available ${z_{{\rm{spec}}}}$. We then calculate the wavelength of the H$\alpha$ line at this redshift and multiply the H$\alpha$ flux with the relative transmission of the NB2300 filter at this wavelength. Next, in order to obtain the narrowband flux of each simulated source, we adopt the dust attenuation-stellar mass relation \citep{Garn2010,Shapley2022} with a 0.4 dex scatter, the contamination from $[{N_{{\rm{II}}}}]$ \citep{Shimakawa2018B}, and a color-term variation 0.1.  Once the ${K_{\rm{s}}}$-band and narrowband fluxes of every simulated source are assigned, we can follow the steps from the last paragraph to test and fit the recovery rate at each stellar mass. The best-fit result is also shown in Figure~\ref{Fig:SC} (upper) and can be written as
\begin{equation}
\label{eq_A3}
C{({\rm{mass}})} = 0.473 \times {\rm{Erfc}}[{\rm{1}}.{\rm{519}} \times (9.079 - \log ({\rm{mass}}))].
\end{equation}

As a robustness test, the SMFs of SFGs with two different incompleteness-correction procedures are compared in Figure~\ref{Fig:SMF_2CC}. We find the top-heavy shape of SMF is robust under different incompleteness-correction procedures. These two SMFs are almost identical at the high-mass end (${M_ * } > {10^{10}}{M_ \odot }$) and only show a small difference around the limiting depth (${M_ * } \sim {10^9}{M_ \odot }$) without affecting the top-heavy shape. Since the best-fit parameters in Equation~\ref{eq_A3} might be influenced by the uncertainty (e.g. environmental dependence and the bias from observation) of the stellar mass-${K_{\rm{s}}}$ flux relation, SF main sequence, and dust attenuation, we use Equation~\ref{eq_A1} for the incompleteness correction in Section~\ref{sec: Results}. We caution that the completeness discussed here is targeted at the sample of HAEs. This tends to slightly underestimate the completeness of the full SF sample since we also include four additional non-HAE spectroscopic members (mostly at ${M_ * } = {10^{10.0 - 10.5}}{M_ \odot }$).

\section{The Schechter fitting of SMFs}\label{Appendix_fitting_results}
In order to avoid the uncertainty caused by the arbitrary binning procedure, the SMFs are fitted with maximum-likelihood estimation (MLE). We perform the MLE fitting by minimizing the negative log likelihood \citep[e.g.][]{Marshall1983,Hill2022}, which can be written as
\begin{equation}
 - \ln L = \int_{{M_{{\rm{*,min}}}}}^{{M_{{\rm{*,max}}}}} {\Phi ({M_{\rm{*}}})} {\rm{d(}}{M_{\rm{*}}}) - \sum\limits_{i = 1}^N {\ln } \Phi ({M_{{\rm{*,}}i}})/C({\rm{NB}})
\end{equation}
where ${M_{{\rm{*,min}}}} = {10^{9.2}}{M_ \odot }$ and ${M_{{\rm{*,max}}}} = {10^{12.0}}{M_ \odot }$ are the lower and upper limits of the fitting range, respectively, and $C({\rm{NB}})$ is the incompleteness-correction factor given by Equation~\ref{eq_A1} in Appendix~\ref{Appendix_completeness}. During the fitting, since the double Schechter function can be considered as the mixture of two single Schechter functions, which is similar to the mixture of gamma distributions \citep{Young2019}, we use the Expectation Maximization algorithm \citep{Dempster1977} to determine the contribution of each member to each Schechter component. In addition, when we are fitting the SMF of SFGs, the power-law slope of the double Schechter function is fixed at ${\alpha _1} =  - 1.5$ \citep[e.g.][]{Shimakawa2018A,Shimakawa2018B} because it cannot be well constrained by our limited mass range and sample size. To obtain the uncertainty of the best-fit parameters, we randomly determine the stellar mass of each galaxy according to the mass uncertainty given by Bagpipes, as well as its weight considering the Poisson error. We repeat this procedure to perform the MLE for 100 times and then take the standard deviation of the 100 best-fit values as the uncertainty. All of the best-fit parameters and their uncertainties are listed in Table~\ref{Tab:SMF_parameters}.

\begin{table*}[!tbh]\small
\centering
\begin{minipage}[center]{\textwidth}
\centering
\caption{Best-fit parameters of the stellar mass functions. \label{Tab:SMF_parameters}}
\begin{tabular}{lcccccc}
\hline\hline
  & $\log ({M^ * })$  & $\Phi _{_1}$ & ${\alpha _1}$ &  $\Phi _{_2}$ & ${\alpha _2}$ & $\Delta ({\rm{BIC)}}$$^a$ \\
  & $({M_ \odot })$& $({\rm{de}}{{\rm{x}}^{ - 1}})$ & & $({\rm{de}}{{\rm{x}}^{ - 1}})$ & &  \\
\hline
SFGs &  $10.52 \pm 0.14$ & $4.00 \pm 0.58$ & [-1.5]$^b$ &$31.7 \pm 3.4$ &$0.79 \pm 0.58$ & 11.02\\
SFGs (HAEs only) &  $10.42 \pm 0.12$ & $4.79 \pm 0.53$ & [-1.5] &$20.4 \pm 3.1$ &$1.38 \pm 0.55$ & 10.03\\
SFGs (merged) &  $10.44 \pm 0.11$ & [4.00] & [-1.5] & $24.6 \pm 1.6$ & $0.40 \pm 0.34$ & 1.01 \\
SFGs (after quenching) &  $10.41 \pm 0.10$ & [4.00] & [-1.5] & $21.4 \pm 1.3$ & $0.54 \pm 0.37$ & 0.96\\
QGs  &  $10.28 \pm 0.20$ & $4.3 \pm 1.0$ & $1.61 \pm 0.89$ & & & -2.21\\
QGs (after quenching) &  $10.58 \pm 0.16$ & $21.1 \pm 1.7$ & $0.78 \pm 0.36$ & & & -0.91\\
\hline
\end{tabular}
\begin{flushleft}
{\sc Note.} --- 
($a$) $\Delta ({\rm{BIC)}} \equiv {\rm{BI}}{{\rm{C}}_{{\rm{single}}}} - {\rm{BI}}{{\rm{C}}_{{\rm{double}}}}$, $\Delta ({\rm{BIC)}} > 0$ means that the double Schechter function can describe the sample better than the single Schechter function, and the typical threshold for a valid comparison is $\Delta ({\rm{BIC)}} > 8 \sim 10$ \citep{Kass1995}.
($b$) Fixed parameters are enclosed in square brackets.
\end{flushleft}
\end{minipage}
\end{table*}

\bibliography{J1001_reference}
\bibliographystyle{aasjournal}

\end{document}